\documentclass[times,twocolumn,final,authoryear]{elsarticle}

\usepackage{ycviu}
\usepackage{framed,multirow}

\usepackage{amssymb}
\usepackage{latexsym}

\usepackage{url}
\usepackage{xcolor}

\usepackage{multirow}
\usepackage{algorithm}
\usepackage{algorithmicx}
\usepackage{algpseudocode}
\usepackage{amsmath}
\usepackage{color}

\definecolor{newcolor}{rgb}{.8,.349,.1}

\journal{Computer Vision and Image Understanding}

\begin{document}

\thispagestyle{empty}

\clearpage
\thispagestyle{empty}
\ifpreprint
  \vspace*{-1pc}
\fi

\clearpage
\thispagestyle{empty}

\ifpreprint
  \vspace*{-1pc}
\else
\fi

\clearpage

\ifpreprint
  \setcounter{page}{1}
\else
  \setcounter{page}{1}
\fi

\begin{frontmatter}

\title{Brain Tumor Anomaly Detection Via Latent Regularized Adversarial Network}

\author[1]{Nan \snm{Wang}}
\author[1]{Chengwei \snm{Chen}}
\author[1]{Yuan  \snm{Xie}}
\author[1]{Lizhuang \snm{Ma}}
\cortext[cor1]{Corresponding author 
  }
\ead{lzma@cs.ecnu.edu.cn}

\address[1]{East China Normal University, North Zhongshan Road Campus: 3663 N, Shanghai and 200062, China}



\received{}
\finalform{}
\accepted{  }
\availableonline{  }
\communicated{}

\begin{abstract}
With the development of medical imaging technology, medical images have become an important basis for doctors to diagnose patients. The brain structure in the collected data is complicated, thence, doctors are required to spend plentiful energy when diagnosing brain abnormalities. Aiming at the imbalance of brain tumor data and the rare amount of labeled data, we propose an innovative brain tumor abnormality detection algorithm. The semi-supervised anomaly detection model is proposed in which only healthy (normal) brain images are trained. Model capture the common pattern of the normal images in the training process and detect anomalies based on the reconstruction error of latent space. Furthermore, the method first uses singular value to constrain the latent space and jointly optimizes the image space through multiple loss functions, which make normal samples and abnormal samples more separable in the feature-level. This paper utilizes BraTS, HCP, MNIST, and CIFAR-10 datasets to comprehensively evaluate the effectiveness and practicability. Extensive experiments on intra- and cross-dataset tests prove that our semi-supervised method achieves outperforms or comparable results to state-of-the-art supervised techniques.
\end{abstract}

\begin{keyword}
\MSC 41A05\sep 41A10\sep 65D05\sep 65D17
\KWD latent regularized\sep Brain tumor detection\sep adversarial network\sep Singular Value Decomposition

\end{keyword}

\end{frontmatter}






\section{Introduction}
Brain lesions can be formulated as the tissue abnormalities issues, which are caused by many possible reasons, such as cancer, infection and disease. The early detection plays a critical role in the treatment of most lesions, which could prevent severe symptoms before they arise. In recent years, medical images have become an important basis for doctors in clinical diagnosis, disease tracking and teaching research. Brain lesions could be detected from medical images. Especially, Magnetic Resonance Imaging (MRI), provides the necessary information of the brain structure. With the development of machine learning and imaging technologies, they provide a viable solution to accelerate radiological researches and make detection progress more efficient. However, the lacking of sufficient negative samples from brain lesions dataset, leads to the data imbalance issue existing in the detection process. Some unknown negative samples exist in the dataset. Therefore, the main challenge of brain tumor detection is the way to achieve robustness and generalization to different kinds of brain lesions.

\begin{figure}[h]
	\centering
	\includegraphics[width=\linewidth]{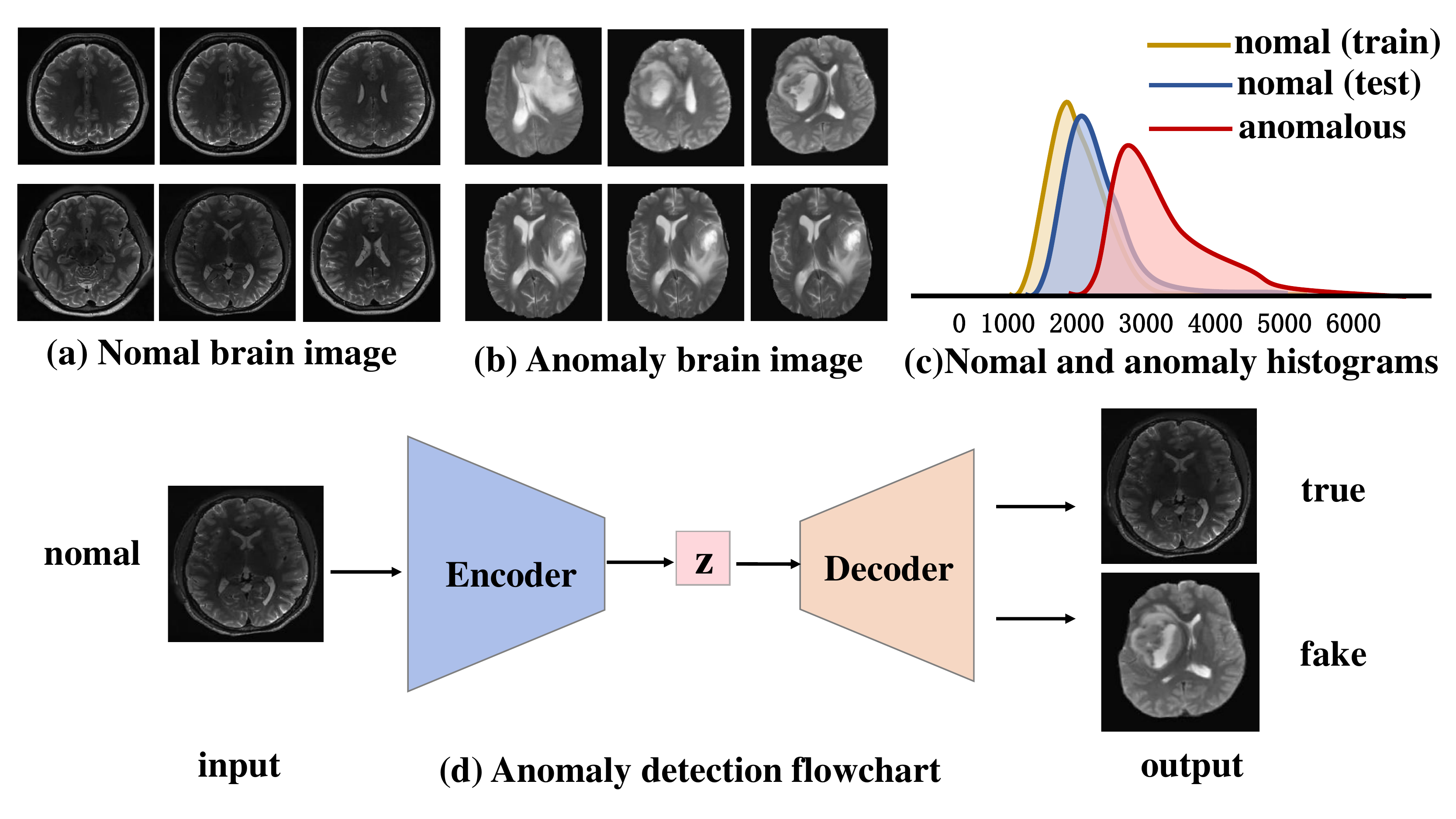}
	\caption{Anomaly detection flowchart}
	\label{introduction}
\end{figure}

Most of previous brain tumor detection work focus on supervised methods, with the utilization of hand-crafted or learned features. Most approaches typically depend on binary supervision. Nevertheless, many previous works suffer from the following major limitations partially or wholly: (1) fully supervised setting--the utilization of both normal and abnormal brain data (with labels), (2) the assumption of binary classification, and (3) the impracticality to take all types of brain lesions (maybe unknown lesion) into account. Furthermore, collecting lesion brain data for training purpose is costly and time-consuming. Binary supervision also could be insufficient to learn a valuable model and make desired predictions in real world complex scenario. As a result, those brain tumor detection techniques have limited robustness and generalization to various types of brain lesions.

Technology of anomaly detection makes progress from binary supervised learning to semi-supervised learning. Given the images of the healthy brain (inliers), an alternative approach is tantamount to model the distribution of healthy brains. Since the parameters of the trained model are more suitable to represent the vigorous brains image well, the images with brain lesions (outliers) could not be represented well with high reconstruction error. Thanks to a supervised training signal coming from the reconstruction objective, autoencoders methods \citep{an2015variational,kingma2013auto,ng2011sparse} could capture the modes from normal samples and distinguish the normal and abnormal sample by reconstruction error. Unfortunately, they suffer from memorization and have a tendency to produce blurry images. The original intention of autoencoder methods is dimensionality reduction, which could not capture the full real concept of target class, leading to the blurry image reconstruction. It is hard to distinguish normal from abnormal samples by reconstruction error.

In recent, GANs \citep{goodfellow2014,schlegl2017unsupervised,chen2018unsupervised} have been successfully adopted in abnormal detection. GANs catch the distribution of the target class by taking a two-player minimax game and usually consists of the generator and discriminator. The generator attempts to generate realistic image to deceive the discriminator. The discriminator attempts to distinguish the generated image from the original image. Both of them compete with each other while capture the underlying concept from the target class.
During inference, test samples from the learned distribution, should get a fair representation in the latent space and anomalous samples will not. However, previous GAN methods focus on the optimization of image space. The abnormal samples are identified by pixel-wise reconstruction error. The image reconstruction error is not strongly related to the abnormal detection task. The aim of this task is to determine the difference between positive and negative samples in the latent feature space.

Motivated by the above limitations, we formulate the brain tumor detection as a anomaly detection task, due to the following reasons: (1) Generalization issue could be tackled by adopting the methodology of anomaly detection, where the attacks are regarded as the out-of-distributions samples that naturally exhibit a higher reconstruction error in the latent space than real samples. (2) It is impractical to take all types of brain lesion (maybe unknown lesion) into account, meanwhile collecting brain lesion image training purpose is costly and time-consuming in the supervised learning. Therefore, we design a GAN style anomaly detection framework which could learn the latent space for the target class and detect the outlier anomalies (brain lesion images) by only training the model for the normal class (healthy brain images).

\begin{figure*}[!h]
	\centering
	\includegraphics[width=\linewidth]{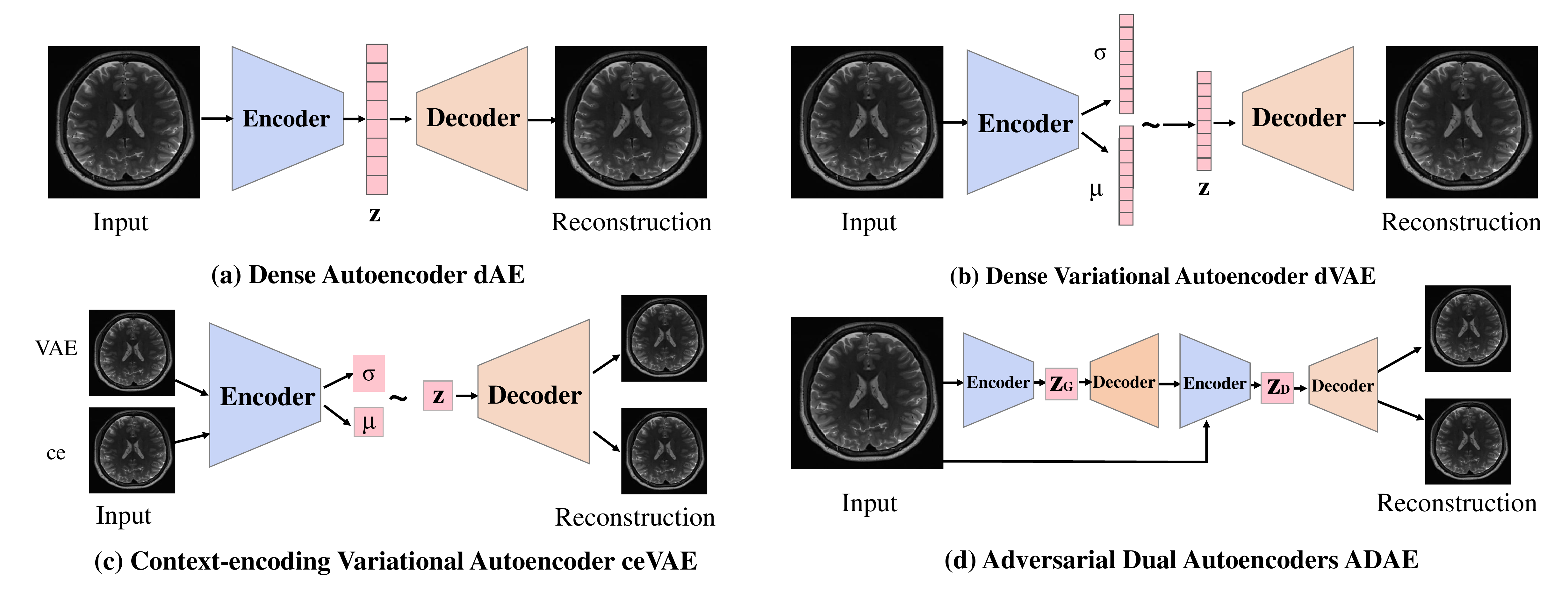}
	\caption{ An overview of diﬀerent Autoencoder frameworks}
	\label{comparefram}
\end{figure*}

In this paper, to obtain a more accurate abnormal detector, we focus on the optimization of latent feature space, apart from image space. Fig. \ref{introduction} is the schematic diagram, the algorithm only models the normal samples during training, therefore, the abnormality is judged by the reconstruction difference of the test samples. A novel resonant network for brain tumor detection has been proposed by using adversarial training under semi-supervised learning framework. Compared with conventional GAN style architecture, we add two components, including a latent regularizer and an auxiliary encoder. Apart from the reconstruction error, the latent regularizer is further utilized to distinguish between normal samples and outliers in latent feature space. The auxiliary encoder $\left(G_{e^{\prime}}(\cdot)\right)$ is adopted to minimize the distance between the bottleneck features of the original input image and encoded generated image latent feature, it acts as an anchor point to prevent from drifting in latent feature space during model optimization, resulting in a more concentrated distribution for the target class.

\begin{itemize}

	\item 
	 A novel decoder-encoder-decoder framework increases the generator capacity to capture the real concept of target class under adversarial optimization schemes.

	\item   The underlying structure of training data is not only captured in image reconstruction space, but also can be further restricted in the space of latent representation in a discriminant manner, leading to a more robust detector.

	\item   We conduct the extensive evaluation of our semi-supervised method on several challenging datasets, where the experimental results demonstrate that our method outperforms or comparable result to  state-of-the-art competitors. 

\end{itemize}

\section{Related Work}
\subsection{Deep autoencoders in Anomaly Detection}

The deep autoencoders \citep{hinton2006reducing} are the primary method used for deep anomaly detection, autoencoders are the neural networks that attempt to learn the identity function and have an intermediate representation of dimensionality reduction, which is the bottleneck that causes the network to extract salient features from certain datasets. The objective of these networks is tantamount to minimizing the reconstruction error between the original image and the reconstruction image. Hence, deep autoencoders based method capture the common factors of variation from normal samples and represent target class well. The out-of-distribution samples without these universal factors of variation and obtain high reconstruction error. Therefore, the reconstruction error is regarded as abnormal score to distinguish the normal samples from abnormal samples. Some variants of the autoencoder are proposed to the abnormal detection task, such as denoising autoencoders \citep{vincent2010stacked}, variational autoencoders (VAEs)\citep{kingma2013auto} and deep convolutional autoencoders (DCAEs) \citep{makhzani2015winner}.

 
Along with their obvious advantages, deep autoencoders have critical disadvantages. The original objective of deep autoencoders methods is dimensionality reduction. The biggest challenge of deep autoencoders method is to choose the right degree of compression. If there is no compression in the network, the autoencoder just learn the identity function. If the input data is compressed to a single value, the mean of input data is the optimal solution. The “compactness” of the image representation is a model hyperparameter. It is laborious for us to find the proper balance between those two cases.

\subsection{Generative Adversarial Networks in Anomaly Detection}

Apart from autoencoders, GANs have been successfully adopted in abnormal detection task. Only normal samples regarded as the target class is trained in GANs. The GANs learn the distribution from the target class under the adversarial learning framework. GANs include generator and discriminator. The generator learns the conventional factors of variation from a normal class and produces fake images to deceive discriminator. The objective of discriminator is tantamount to distinguish the inliers from outliers. The generator and discriminator compete with each other while capturing the underlying characteristics from training normal data.

During the inference, since the parameters of the learned model are more suitable to reconstruct the normal samples, the out-of-distribution samples obtain high reconstruction error. The discriminator’s prediction in the image space is utilized to quantify the reconstruction error. The discriminator output the probability of novelty class. EGBAD \citep{zenati2018efficient} uses a Bidirectional GAN (BiGAN) \citep{donahue2016} to learn an encoder, which is the inverse of a generator that maps image space to latent space, then, combine the reconstruction loss and the discriminator-based loss to calculate the abnormal score. Schlegletal \citep{schlegl2017unsupervised} recently proposed a novel deep anomaly detection method based on GANs called AnoGAN. AnoGAN attempts to find the point closest to the test input sample in the latent space of the generator. Intuitively, if the GAN captures the distribution of the training data, then the normal samples should have a good representation in the latent space, while the anomalous samples will not, AnoGAN ultimately defines the anomaly score through reconstruction errors. However, all of previous GANs methods only consider the optimization of image space. The objective of abnormal detection task is to identify more separable features between normal and abnormal samples. Therefore, the latent regularizer is proposed to optimize the latent feature space of the target class.

\begin{figure*}[h]
	\centering
	\includegraphics[width=\linewidth]{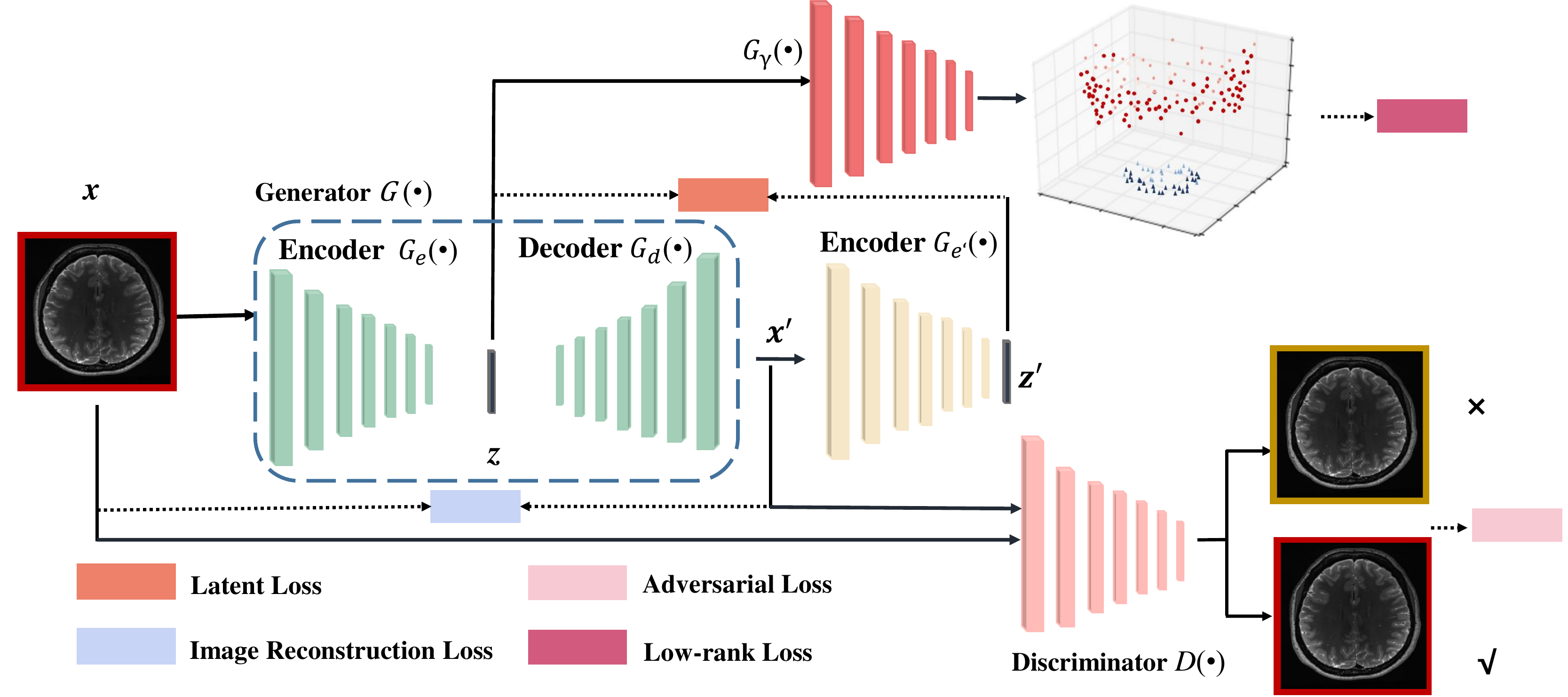}
	\caption{ Our framework consists of a generator, a discriminator, a latent regularizer and an auxiliary encoder. The generator and discriminator were trained by competing with each other while collaborating to understand the underlying concept in the normal class. The latent regularizer is further utilized to distinguish between normal samples and outliers in a discriminant way in latent feature space. The auxiliary encoder is adopted to minimize the distance between the bottleneck features of the original input image and encoded latent feature of the generated image.}
	\label{framework}
\end{figure*}

\subsection{Semi-supervised and Unsupervised Brain Tumor Detection}

In the field of brain tumor detection, regarding tumor detection network (layer-by-layer detection), was a feed-forward neural network with multiple latent layers. These networks seek to reconstruct the input data at the output layer. However, training multi-layer autoencoders is tedious. This is because the weights of deep latent layers are strait to optimize.
\cite{pawlowski2018unsupervised} trains the internal brain CT dataset, they consider the pixel-level reconstruction errors of different AE models for pixel-wise anomaly detection. In their evaluation, AE with dropout samples at the bottleneck level is slightly better than other models. Although they are often used, most reconstruction-based methods have no formal assertions about reconstruction errors, which complicate the interpretation and comparability of anomaly scores. \cite{chen2018unsupervised} shows that the combination of VAE and the antagonistic loss of latent variables can boost the performance in BraTS datasets, it can detect pixel tumors using pixel-wise reconstruction errors. \cite{zimmerer2018context} use VAE to form ceVAE context coding, which provided reconstruction errors and internal model variations. The anomaly score is the estimated probability of the test sample collected from the ELBO. \cite{vu2019anomaly} advanced the Adversarial Dual Autoencoders (ADAE) model, which consists of two autoencoders as generators and discriminators to improve the stability of training. 

More relevant to medical image anomaly detection, some recent work, such as \cite{schlegl2017unsupervised} and \cite{Sato2018A} proposes methods for detecting anomalous regions. \cite{schlegl2017unsupervised} proposed the AnoGAN in trains a GAN on retinal optical coherence tomography images, and then for a given test image, the corresponding “healthy” image is determined by the gradient descent during the latency period. Then, the difference between the reconstructed image and the original image is used to determine the abnormal score of the entire image, and the pixel-level difference is utilized to detect the anomalous region in the image. This method differs from VAE-based methods in that they reconstruct the "healthy" version of the input image. \cite{Sato2018A} applied an unsupervised anomaly detection method to detect lesion of emergency head CT, in the training stage, they used normal 3D patches to train 3D convolutional autoencoder, in the test phase, they calculated the abnormality (normal or abnormal) of each voxel in emergency head CT, and evaluated the possibility of lesions.

\section{Proposed Approach}

In this section, we present how to learn the intrinsic structure of the target class by using the proposed adversarial training framework. We start by describing the details of the overview network architecture, then depict each term in loss function, and finally give the optimization method.

\subsection{Network Architecture}
The GAN-style architecture, which is shown in Fig. \ref{framework}, consists of four components: the encoder-decoder part, the discriminator, auxiliary encoder, and latent regularizer.

 Generator network $\left(G(\cdot)\right)$ includes an encoder and a decoder. The generator network reconstructs the original input images to fool the discriminator, and the discriminator network $\left(D(\cdot)\right)$ needs to distinguish between real images and reconstructed ones. During training procedure, both sub-networks compete with each other to attend high-quality reconstructions that even the discriminator can be established. The model captures the real distribution of the target class by minimizing the Pixel-wise reconstruction error between the original input image and the reconstructed image.
The parameters of training generator are more suitable to reconstruct normal samples. Therefore, anomalies could not be represented well from training model. The generator sub-network obtains the input image $x$ and passes it through the encoder $\left(G_{e}(\cdot)\right)$, and then compresses it to a latent representation $z$ through the convolution layer to reduce $x$. On the other hand, the main operators of decoder $\left(G_{d}(\cdot)\right)$ include convolutional transpose layer, Relu activation and  Batch Normalization(BN) with the last tanh layer, which is used to enlarge latent feature vector $z$ to reconstruct image $x$. The structure of the discriminator is a series of convolutional layers, in order to prevent being fooled by the generator, the discriminator learns the distribution of the target class in adversarial training. During the training process, the discriminator also assists the generator to obtain more robust and stable parameters under the adversarial scheme. 

For further identify in low dimensional space, low rank constraint is adopted in our model. The singular value of minimal rank constraint is utilized to distinguish normal samples and abnormal samples. Singular Value Decomposition(SVD) is used primarily to reduce the dimension of sample data. We input abundant positive samples to train the network, SVD can extract the common characteristics of the samples, thereby obtaining the general distribution of positive samples. Therefore, the distinction between positive and negative samples becomes relatively uncomplicated.

The auxiliary encoder $\left(G_{e}{\prime}(\cdot)\right)$ is adopted to minimize the distance between the bottleneck features $z$ of the original input image and encoded latent feature $z{\prime}$ of the generated image. The reason why we employ the auxiliary encoder which has the same architecture as $G_{e}$ in generative sub-network but with different parametrization, can be concluded as follows: (1) Only for normal samples, the latent representation $z$ can be well reconstructed by the feature vector encoded from the generated image $x^{\prime}$. (2) Since the latent regularizer might incur the distribution distortion in latent feature space, the feature representation $z{\prime}$ can be regarded as the anchor to prevent $z$ from drifting.

\subsection{Overall Loss Functions}
To train the model, we define a loss function, as shown in formula \ref{total}, it contains four parts, including the image reconstruction loss $L_{irec}$, the adversarial loss $L_{adv}$, the latent representation loss $L_{zre}$ and the low-rank loss $L_{rank}$.

 \begin{equation}
 \label{total}
\mathcal{L}=w_{i} \mathcal{L}_{i r e c}+w_{a} \mathcal{L}_{a d v}+w_{z} \mathcal{L}_{z r e c}+w_{r} \mathcal{L}_{rank}
\end{equation}

Where $w_{i}$, $w_{a}$, $w_{z}$, and $w_{r}$ are the weighting parameters balancing the impact of individual term to the overall objective function.

\textbf{Image reconstruction Loss: } While the discriminator tries to differentiates between authentic samples and generated samples, and the generator trying to fool the discriminator. However, the generator is not optimized towards learning the authentic concept from input data only by adversarial loss. Some prior work \cite{isola2017image} has proposed that the distance between input images and generated images should be considered. Isola \citep{isola2017image}  shows that the use of $L1$ yields less blurry results than $L2$. Therefore, we use $L1$ loss function to penalize the generator by minimizing the distance between original input $x$ and generated images $G(x)$ as follows.

 \begin{equation}
\mathcal{L}_{i r e c}=\mathbb{E}_{x \sim p_{\mathbf{x}}}\|x-G(x)\|_{1}
\end{equation}

\textbf{Adversarial Loss:} Adversarial loss is applied to train two networks, including generative sub-network and discriminator $D$ .  \cite{goodfellow2014} proposed a two-player min-max game: $D$ is learned to distinguish generated images from real images, while the generator $G$ is trained to fool $D$. This adversarial game between the generator and discriminator can be formulated as:
\begin{equation}
\begin{aligned}
\mathcal{L}_{a d v} &=\min _{G} \max _{D}\left(E_{\boldsymbol{x} \sim p_{\mathbf{x}}}[\log (D(\mathbf{x}))]\right.\\
&\left.+E_{\boldsymbol{x} \sim p_{\mathbf{x}}}[\log (1-D(G(\mathbf{x})))]\right)
\end{aligned}
\end{equation}

 $G$ learns the underlying concept in the normal class and is trained to maximally confuse the $D$ into believing that samples it generates come from the data distribution. $D$ tries to discriminate between actual data and the fake data generated by $G$. The $G$ is formed by leveraging the gradient of $D(x)$, and using that to modify its parameters.

\textbf{Latent representation loss: }
Two loss functions described above can support the generator to learn the real concept in the target class and only consider the distance between original input $x$ and generated images $x^{\prime}$.  \cite{donahue2016} suggests that an encoder network is used to map from image space to latent space, a vanilla GAN network is capable of learning inverse mapping. Only for target class samples, the auxiliary encoder can reconstruct the latent representation $z$ well from generated image $x$. Besides, the latent regularizer might incur the distribution distortion in latent feature space, the feature representation $z^{\prime}$ can be regarded as the anchor to prevent $z$ from drifting. Moreover, A lot of work based on image reconstruction error, \cite{sabokrou2018} have been proposed in anomaly detection task. Using reconstruction error to detect anomalies is not our original intention. We consider to find more separable features between normal sample and abnormal sample by optimizing the latent feature space. Hence, we add a constraint by minimizing the distance between latent feature of generator input images $G_{e}(x)$ and generated image latent feature from auxiliary encoder $G_{e^{\prime}}\left(x^{\prime}\right)$ as follows.

\begin{equation}
\mathcal{L}_{z r e c}=\mathbb{E}_{x \sim p_{\mathbf{X}}}\left\|G_{e}(x)-G_{e^{\prime}}\left(x^{\prime}\right)\right\|_{2}
\end{equation}

\textbf{low-rank Loss: } In order to distinguish the data in the low-dimensional space $z$, the latent space should be constrained. SVD is the way of low-rank approximation, which is a kind of discriminative model to be widely used for outlier detection. Given a generator $G_{\gamma}$, measurement sequence $y_{t}$ and measurement matrix $A_{t}$. We suppose that the variation in the image sequence is localized and the image sequence can be mapped to a low dimension space. For each image, we define a matrix $Z$, $z_{t}$ is the $t^{t h}$ corresponding to the latent sequence of the image. To explore low-rank embedding, we solve the following constraint optimization.

\begin{equation}\begin{array}{l}
\min _{z_{1}, \ldots, z_{T}, \gamma} \sum_{t=1}^{T}\left\|y_{l}-A_{t} G_{\gamma}\left(z_{t}\right)\right\|_{2}^{2} \\
\text { s.t. } \operatorname{rank}(Z)=r
\end{array}\end{equation}

\subsection{Optimization}

The structure of generator sub-network and inference sub-network based on DCGAN\citep{radford2015unsupervised}. During training, Adam-optimizer\citep{kingma2014adam} was used to optimize the parameters in the proposed framework. The learning rate of the network is placed at 0.002. The reconstruction error of latent space is utilized as the abnormal score, we noticed that the anomalous score calculation we chose can achieve a preferable division between normal score and outlier score, resulting in higher performance.

\begin{table*}[h!]
	\scriptsize
	\centering 
	\renewcommand\tabcolsep{11.0pt} 
	\caption{One-class novelty detection results for MNIST dataset}
	\label{mnist}
	\begin{tabular}{c|c|c|c|c|c|c|c|c|c|c|c}
		\hline 
		&\textbf{0} &\textbf{1}&\textbf{2}&\textbf{3}&\textbf{4}&\textbf{5}&\textbf{6}&\textbf{7}&\textbf{8}&\textbf{9}&\textbf{MEAN}\\
		\hline 
		
		\textbf{KDE ('06) }&0.885&0.996&0.710&0.693&0.844&0.776&0.861&0.884&0.669&0.825&0.8143\\

		\textbf{DAE  ('06) }&0.894&\textbf{0.999}&0.792&0.851&0.888&0.819&0.944&0.922&0.740&0.917&0.8766\\

		\textbf{VAE  ('13) ('13)}&0.997&\textbf{0.999}&0.936&0.959&0.973&\textbf{0.964}&\textbf{0.993}&\textbf{0.976}&0.923&0.976&0.9696\\

		\textbf{Pix CNN  ('16)}&0.531&0.995&0.476&0.517&0.739&0.542&0.592&0.789&0.340&0.662&0.6183\\

		\textbf{GAN  ('17)}&0.926&0.995&0.805&0.818&0.823&0.803&0.890&0.898&0.817&0.887&0.8662\\

		\textbf{AND  ('19)}&0.984&0.995&0.947&0.952&0.960&{0.971}&0.991&0.970&0.922&\textbf{0.979}&\textbf{0.9671}\\

		\textbf{AnoGAN  ('17)}&0.966&0.992&0.850&0.887&0.894&0.883&0.947&0.935&0.849&0.924&0.9127\\

		\textbf{DSVDD ('18)}&0.980&0.997&0.917&0.919&0.949&0.885&0.983&0.946&\textbf{0.939}&0.965&0.9480\\

        
        \textbf{IGMM-GAN  ('18)
		}&0.955&0.900&0.930&0.820& 0.830 &0.900&0.930&0.900& 0.780&0.570&0.8520\\
        
		\textbf{Proposed method}&\textbf{0.996} &0.985	&\textbf{0.969}&\textbf{0.964}&\textbf{0.981}&0.940&0.917	&0.942&0.921&	0.697	&0.9310\\
		\hline 
	\end{tabular}	
\end{table*}

\begin{table*}[h!]
	\scriptsize
	\centering 
	\renewcommand\tabcolsep{10.0pt} 
	\caption{One-class novelty detection results for CIFAR-10 dataset}
	\label{cifar}
	\begin{tabular}{c|c|c|c|c|c|c|c|c|c|c|c}
		\hline 
		&\textbf{PLANE} &\textbf{CAR}&\textbf{BIRD}&\textbf{CAT}&\textbf{DEER}&\textbf{DOG}&\textbf{FROG}&\textbf{HORSE}&\textbf{SHIP}&\textbf{TRUCK}&\textbf{MEAN}\\
		\hline 
		\textbf{KDE   ('06) }&0.658&0.520&0.657&0.497&0.727&0.496&0.758&0.564&0.680&0.540&0.6097\\

		\textbf{DAE  ('06) }&0.411&0.478&0.616&0.562&0.728&0.513&0.688&0.497&0.487&0.378&0.5358\\
	
		\textbf{VAE  ('13) }&0.700&0.386&\textbf{0.679}&0.535&0.748&0.523&0.687&0.493&0.696&0.386&0.5833\\

		\textbf{Pix CNN  ('16) }&0.788&0.428&0.617&0.574&0.511&0.571&0.422&0.454&0.715&0.426&0.5506\\
	
		\textbf{GAN  ('17) }&0.708&0.458&0.664&0.510&0.722&0.505&0.707&0.471&0.713&0.458&0.5916\\

		\textbf{AND  ('19) }&0.717&0.494&0.662&0.527&0.736&0.504&0.726&0.560&0.680&0.566&0.6172\\

		\textbf{AnoGAN  ('17) }&0.671&0.547&0.529&0.545&0.651&0.603&0.585&0.625&0.758&0.665&0.6179\\

		\textbf{DSVDD  ('18) }&0.617&0.659&0.508&0.591&0.609&0.657&0.677&0.673&0.759&0.731&0.6481\\

		\textbf{OCGAN  ('19) }&0.757&0.531&0.640&0.620&0.723&0.620&0.723&0.575&0.820&0.554&0.6566 \\

		\textbf{Proposed method}&\textbf{0.889}	&\textbf{0.732}&0.665&\textbf{0.697}&\textbf{0.892}&\textbf{0.753}&	\textbf{0.823}&	\textbf{0.704}	&\textbf{0.931}&	\textbf{0.888}&	\textbf{0.7970}\\
		\hline 
	\end{tabular}	
\end{table*}

\section{Experiments}

In this section, we first present the experimental setting, including datasets, and more details involved. Subsequently, toy experiment and ablation study are conducted to analyze the proposed method in detail.  Finally, We testify experimental availability in BraTS dataset.

\textbf{Datasets:} To ensure the fairness of comparison experiments, we firstly apply the proposed algorithm in two standard databases, MNIST \citep{yannmnist} and CIFAR-10 \citep{krizhevsky2009learning}, then compared the results with state-of-the-art methods. Subsequently, our model was trained on the normal brain datasets and applied to BraTS 2019 \citep{menze2014multimodal} for solving real problems. The normal brain data came from the Human Connectome Project (HCP) \citep{van2012human} datasets.

\textbf{MNIST and CIFAR-10: }One of classes would be regarded as an anomaly, while the rest ones belong to the normal class. In total, we get ten sets for MNIST dataset and CIFAR-10 dataset, respectively, then detect the outlier anomalies by only training the model for the normal class.

\textbf{HCP and BraTS 2019: }We use 65 healthy patients in the HCP database as training data and the entire BraTS 2019 dataset as the test set, the BraTS 2019 dataset contains 259 high-grade glioblastomas(HGG) patients and 76 low-grade glioblastomas(LGG), and released data in three subsets of training, verification and testing, including T1, T2, T1ce and Flair in four multiple modalities, and the annotation files are only provided for the training set. The intensity of the image is normalized by z-score normalization to reduce the intensity change of different subjects.

\textbf{Implementation Detail: } Training is done with batch size of $64$, We implement our approach in PyTorch by optimizing the weighted loss $L$ with the weight values $w_{i}=1$,$w_{a}=5$,  $w_{z}=1$, and $w_{o}=0.05$, which are empirically chosen to yield optimum results. The experiments are carried out on a double NVIDIA 2080Ti GPU.

\subsection{Toy experiment}
\label{sec:toy}

In this subsection, we present that the proposed method has the clear superiority over cutting edge semi-supervised abnormal detectors. For MNIST and CIFAR-10 dataset, we select one class as the normal each time, while leaving the rest to be as the anomaly classes, leading to ten sets for abnormal detection. Our method has clear superiority over cutting edge semi-supervised abnormal detectors in the MNIST dataset, as showed in Tab. \ref{mnist}, that is compared with KDE \citep{bishop2006pattern}, DAE \citep{hadsell2006dimensionality}, VAE \citep{kingma2013auto}, Pix CNN \citep{OordConditional}, GAN \citep{schlegl2017unsupervised}, AND \citep{abati2019latent}, AnoGAN \citep{schlegl2017unsupervised}, DSVDD \citep{ruff2018deep} and IGMM-GAN \citep{gray2018coupled}. Compared with IGMM-GAN, which deals specifically with multimodal datasets, our method has achieved significant improvements. For CIFAR-10 dataset, Tab. \ref{cifar} describe the performance of this algorithm compared with  KDE \citep{bishop2006pattern}, DAE \citep{hadsell2006dimensionality}, VAE \citep{kingma2013auto}, Pix CNN \citep{OordConditional}, GAN \citep{schlegl2017unsupervised}, AND \citep{abati2019latent}, AnoGAN \citep{schlegl2017unsupervised}, DSVDD \citep{ruff2018deep} and OCGAN \citep{perera2019ocgan}, the experimental results show that our algorithm is better than other algorithms in most categories, and the average AUC is about $0.797$.

\subsection{Ablation Study}

\begin{table}[h]
	\scriptsize
	\centering 
	\renewcommand\tabcolsep{10.0pt} 
	\caption{The effects of different loss function combinations in toy experimental results}
	\label{loss}
	\begin{tabular}{c|c|c}
		\hline 
		&\textbf{MNIST} &\textbf{CIFAR10} \\
		\hline
		\textbf{$\mathcal{L}_{i r e c}+\mathcal{L}_{a d v}$  }&0.606&0.514\\
		
		\textbf{$\mathcal{L}_{i r e c}+\mathcal{L}_{a d v}+\mathcal{L}_{rank}$ }&0.76&0.532\\

		\textbf{$\mathcal{L}_{i r e c}+\mathcal{L}_{a d v}+\mathcal{L}_{z r e c}$  }&0.85&0.478\\
	
		\textbf{$\mathcal{L}_{i r e c}+\mathcal{L}_{a d v}+\mathcal{L}_{z r e c}+\mathcal{L}_{rank}$  }&0.937&0.797\\
		\hline 
	\end{tabular}	
\end{table}

\begin{table}[h]
	\scriptsize
	\centering 
	\renewcommand\tabcolsep{10.0pt} 
	\caption{The different loss function combinations in BraTS experimental results}
	\label{bratsloss}
	\begin{tabular}{c|c}
		\hline 
		&\textbf{BraTS} \\
		\hline
		\textbf{$\mathcal{L}_{i r e c}+\mathcal{L}_{a d v}$  }&0.027\\
		
		\textbf{$\mathcal{L}_{i r e c}+\mathcal{L}_{a d v}+\mathcal{L}_{rank}$ }&0.684\\

		\textbf{$\mathcal{L}_{i r e c}+\mathcal{L}_{a d v}+\mathcal{L}_{z r e c}$  }&0.835\\
	
		\textbf{$\mathcal{L}_{i r e c}+\mathcal{L}_{a d v}+\mathcal{L}_{z r e c}+\mathcal{L}_{rank}$  }&0.994\\
		\hline 
	\end{tabular}	
\end{table}

The architecture proposed is built on the adversarial training framework. In this paper, a combination of multiple loss functions is utilized to verify the effectiveness, so ablation research is needed. The ablation experiments perform different combined experiments to prove the corresponding performance. For the effectiveness of each component, we conducted relevant studies using the MNIST, CIFAR-10 and BraTS. Specifically, we use four separate function combination modes,the specific experimental results are shown in Tab. \ref{loss} and Tab. \ref{bratsloss}.

In the first case, we detach the latent space reconstruction error loss function and the latent space low rank constraint function simultaneously, the experimental results show that the performance degradation evidently.
In the second case, the latent space reconstruction error function is removed, the results of MNIST experiments decreased by $19\%$ and the performance of the CIFAR-10 underlying network is reduced by $33.2\%$. the BraTS descent rate is $31.2\%$. To further verify the effectiveness of the loss function proposed, in the third case, we remove the low rank constraint function of the latent space, the performance of the network drops significantly, the MNIST result descend rate is $9.3\%$ and the decline rate of CIFAR-10 experimental data is $40\%$. The BraTS experimental results dropped significantly. In the final circumstances, we consider the complete function model, AUC has obtained the best effect for BraTS, MNIST and CIFAR-10.

  At present, most algorithms directly use image space reconstruction errors to calculate anomaly score, they only concentrate on the relevant information at the image level, thereby ignoring the correlation of image pixels in space. While emphasizing the image plane information, our algorithm paid more attention to the associated information represented in latent space. To further understand the importance of external constraints, we compare the performance of latent space to make an intuitive explanation. As showed in Tab \ref{anomaly}, in the BraTS dataset, we compare different types of exception constraint score methods. When using pixel level $(\mathbf{x}-\mathbf{x}^{\prime})$ features to calculate the abnormal score, the result is $0.938$, and the potential characteristics of latent space are used, AUC is $0.994$. Obviously, latent space features $(\mathbf{z}-\mathbf{z}^{\prime})$ performance is enhanced than pixel-level $(\mathbf{x}-\mathbf{x}^{\prime})$ feature. Tab. \ref{anomaly1} presents that the capability of our algorithm is better than conventional anomaly detectors on the MNIST and CIFAR-10. For the MNIST dataset, the anomaly score using latent features is improved by $7.3\%$ compared with the conventional anomaly detection. We also presented the comparison of the CIFAR-10 dataset, Tab. \ref{anomaly1} describes the final impact of the selection of calculating the anomaly score on the overall model performance.

 \begin{table}[t]
	\scriptsize
	\centering 
	\renewcommand\tabcolsep{10.0pt} 
	\caption{The AUC of different anomaly score in BraTS experimental results}
	\label{anomaly}
	\begin{tabular}{c|c}
		\hline 
		&\textbf{BraTS} \\
		\hline
		\textbf{$\mathbf{x}-\mathbf{x}^{\prime}$ }&0.938\\
		
		\textbf{$\mathbf{z}-\mathbf{z}^{\prime}$ }&0.994\\
		\hline 
	\end{tabular}	
\end{table}

 \begin{table}[t]
	\scriptsize
	\centering 
	\renewcommand\tabcolsep{10.0pt} 
	\caption{The AUC of different anomaly score in MNIST and CIFAR-10 experimental results}
	\label{anomaly1}
	\begin{tabular}{c|c|c}
		\hline 
		&\textbf{MNIST}&\textbf{CIFAR} \\
		\hline
		\textbf{$\mathbf{x}-\mathbf{x}^{\prime}$ }&0.868&0.617\\
		
		\textbf{$\mathbf{z}-\mathbf{z}^{\prime}$ }&0.937&0.797\\
		\hline 
	\end{tabular}	
\end{table}

\begin{figure}[h]
	\centering
	\includegraphics[width=\linewidth]{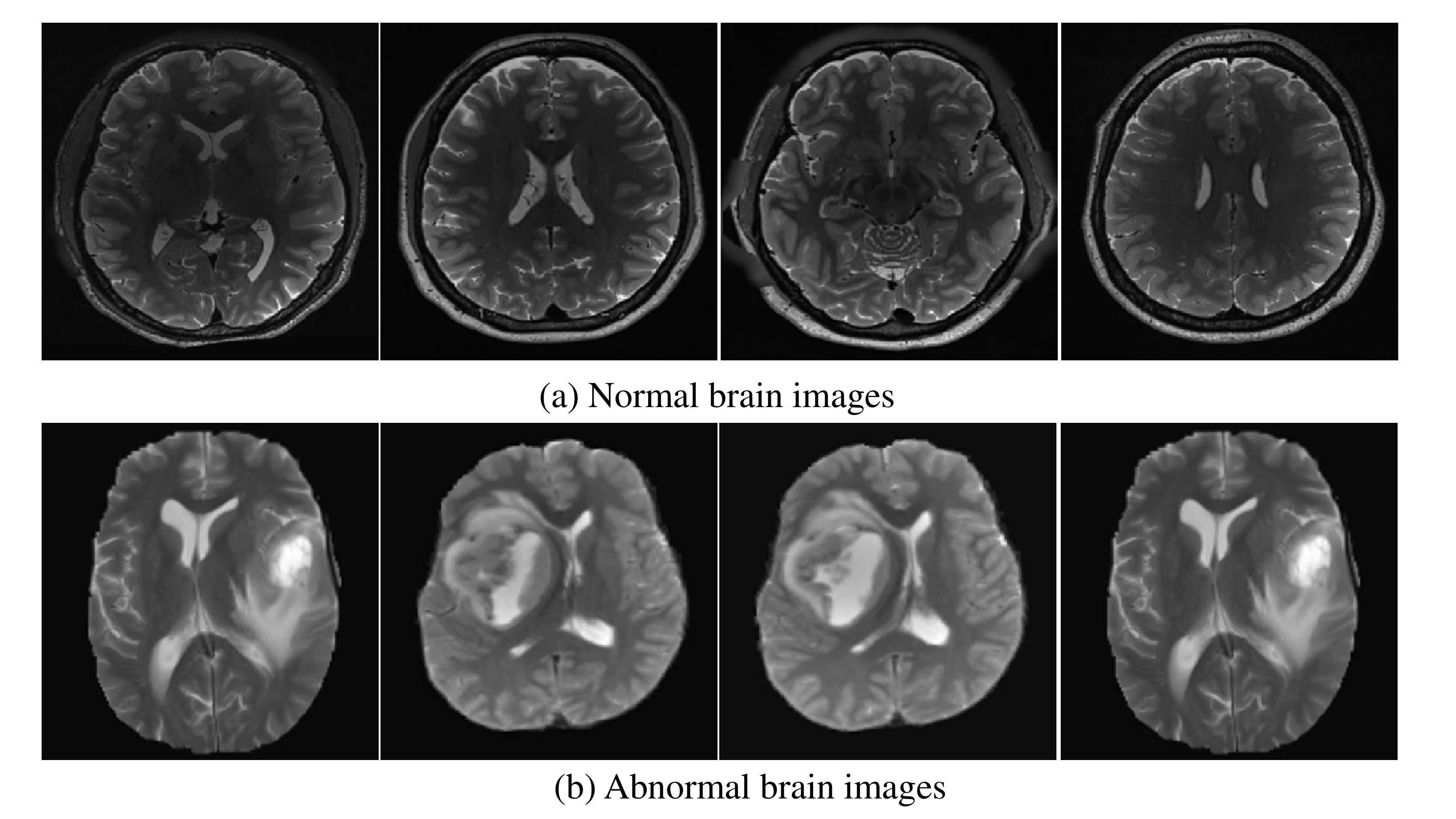}
	\caption{(a)HCP dataset images, (b)BraTS dataset images}
	\label{brats}
\end{figure}

\begin{figure}[h]
	\centering
	\includegraphics[width=\linewidth]{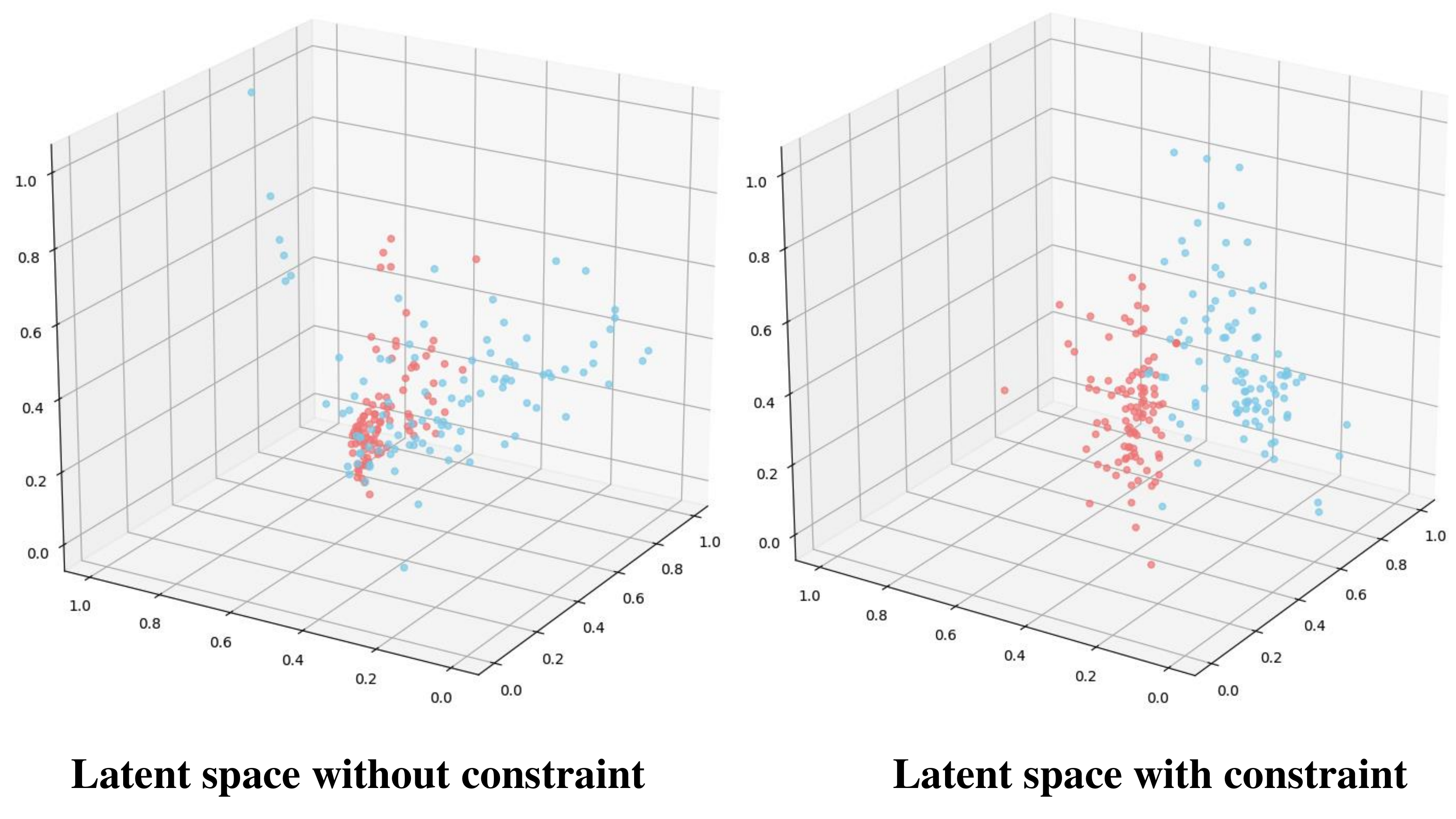}
	\caption{The visualization of latent space learned with and without constraint}
	\label{latent}
\end{figure}

\subsection{Experimental of BraTS dataset}

The types of brain images in dataset are highly inequitable with complex structure, Fig. \ref{brats} shows the data on HCP and BraTS datasets, we selected normal brain images and malformed brain images with obvious divergences. As can be seen from it, (a) are normal brain images, and (b) are abnormal brain images. We only use the normal brain image for training. During the inference, the trained model distinguishes the anomalous samples from normal samples by reconstruction error.

\begin{figure}[h]
	\centering
	\includegraphics[width=\linewidth]{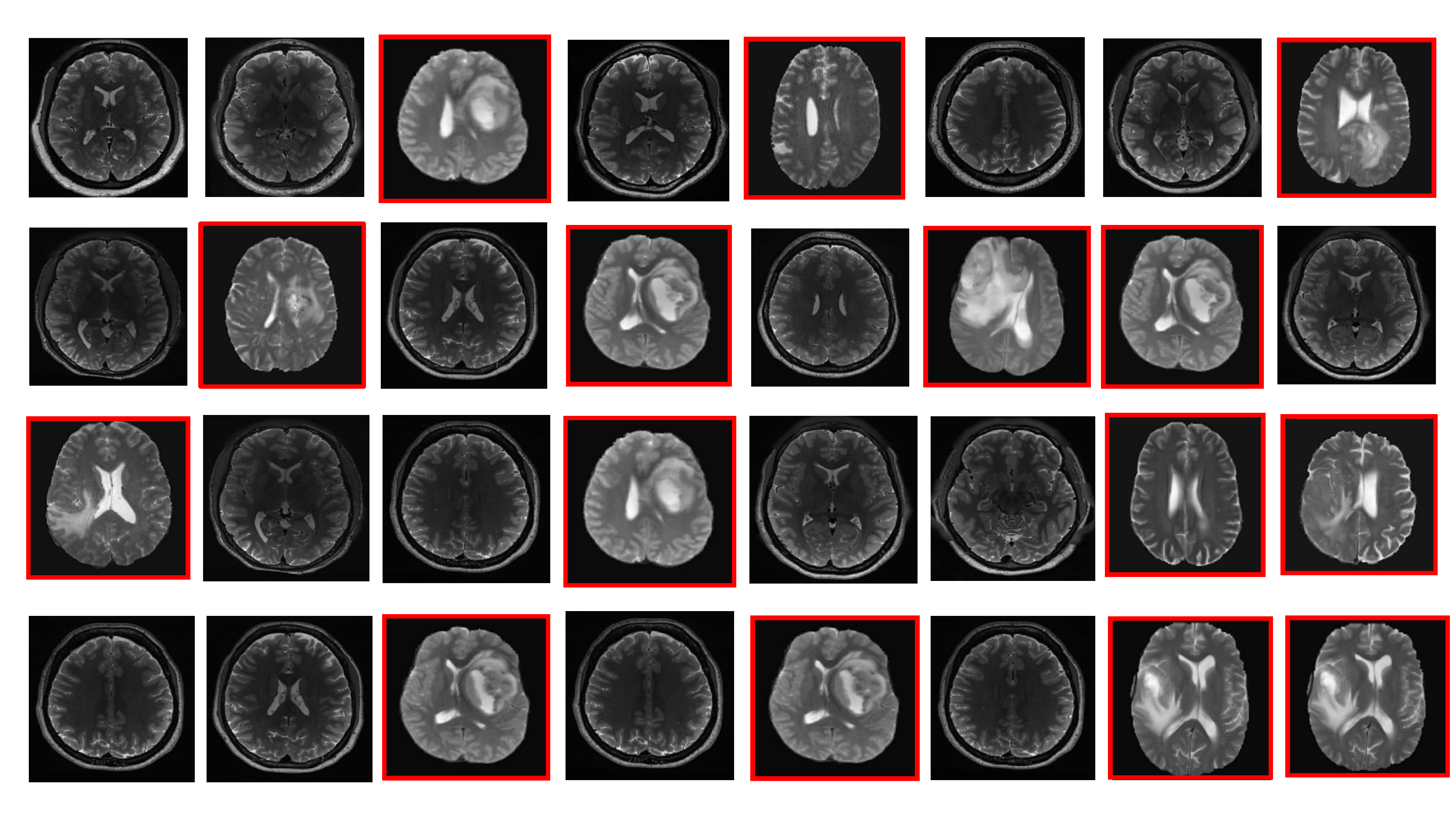}
	\caption{Most anomalous brain in BraTS datasets are detected by proposed method. Abnormal examples are highlighted in red.}
	\label{inputoutput}
\end{figure}

\begin{figure}[H]
	\centering
	\includegraphics[width=\linewidth]{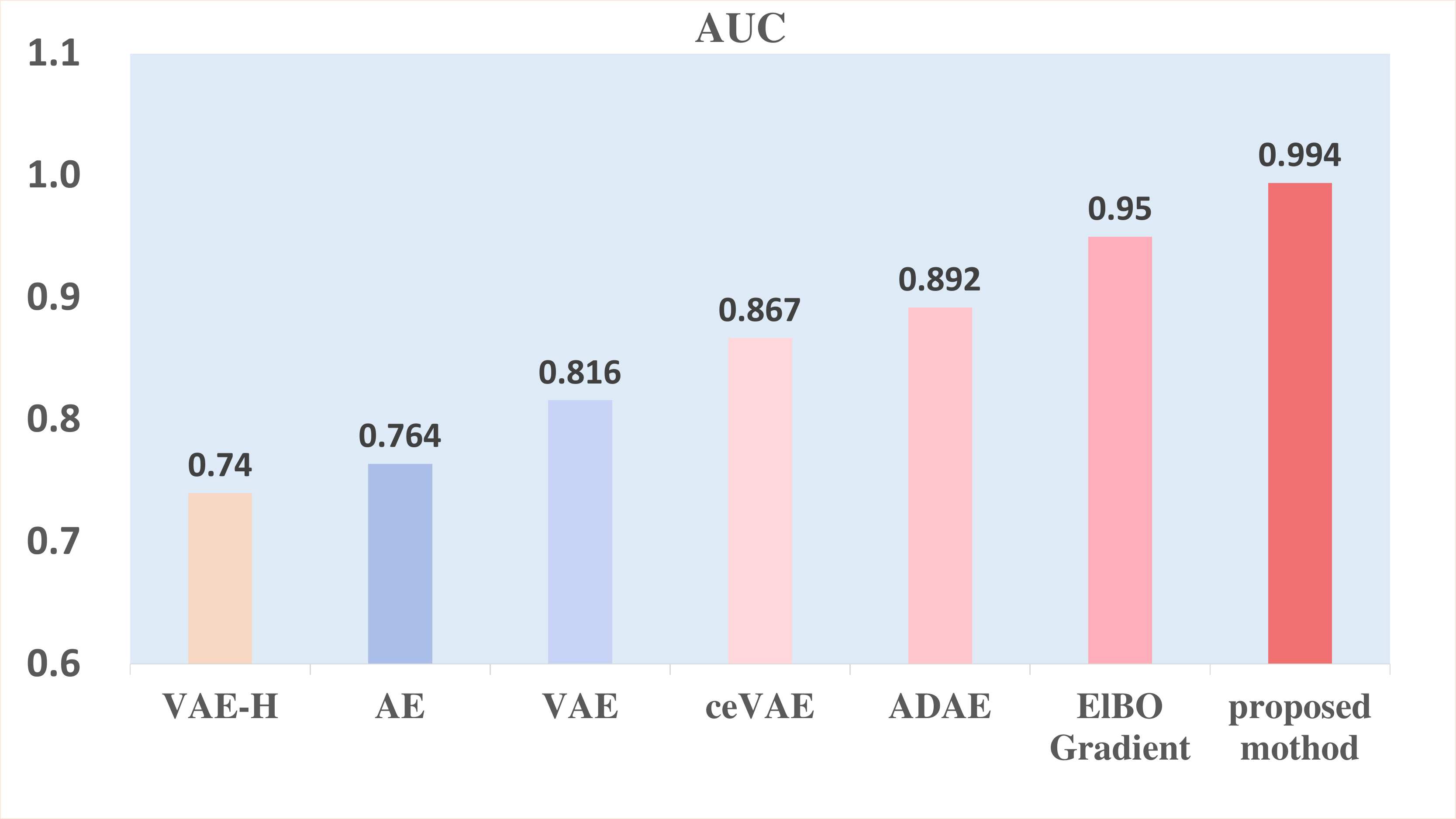}
	\caption{AUC score comparison for BraTS dataset}
	\label{aucshuban}
\end{figure}

\textbf{Latent space optimization evaluation:} Compared with the previous detection strategy focusing on image space optimization, we use multiple losses to jointly optimize the latent space and image space. As we can see in Fig. \ref{latent}, this section gives the evaluation of latent space optimization. In the BraTS experiment, 3D scatter plots are utilized to display the potential features of normal samples (blue dots) and anomalous samples (red dots). It is obvious that the constraint in the latent space makes the normal samples and abnormal samples more separable, which help the model to distinguish the abnormal samples from normal samples.

\textbf{Evaluation Measures:} we employed the performance metric is Area Under Curve(AUC), As shown in Fig. \ref{aucshuban}, our model outperforms previous methods in slice-wise tumor detection such as ceVAE (\cite{zimmerer2018context}), VAE (\cite{baur2018deep}), AE (\cite{chen2018unsupervised}), VAE-H (\cite{albu2020tumor}) and  ELBO Gradient (\cite{zimmerer2019case}).  Fig. \ref{inputoutput} present some test samples detected by the proposed method. The result in Fig. \ref{inputoutput} shows that the most anomalous samples detected by proposed method which includes different types of abnormal samples.

\section{Conclusion}
In order to address the problem of lacking all types of brain lesions (maybe unknown lesion) and optimization of latent space, this paper proposes an adversarial network based on latent space constraints to detect brain tumor abnormalities. During the inference, discriminative reconstruction error of latent space is used as abnormal score, it can effectively distinguish the abnormal samples and normal samples. Experiments demonstrate that the performance of the model is robust in actual brain tumor detection examples such as the HCP and BraTS datasets and two benchmark datasets including the MNIST and CIFAR-10.
\bibliographystyle{model2-names}
\bibliography{refs}

\end{document}